\documentclass[aps,prc,fleqn,floatfix,twocolumn,superscriptaddress,twoside]{revtex4}
\usepackage[]{graphicx}

\usepackage{amsmath,amssymb,amsfonts}

\usepackage{color}

\begin{document}

\title{Elliptic and triangular flow anisotropy in  deuteron-gold collisions at RHIC and proton-lead collisions at the LHC}

\author{Guang-You Qin}
\affiliation{Institute of Particle Physics and Key Laboratory of Quark and Lepton Physics (MOE), Central China Normal University, Wuhan, 430079, China}
\affiliation{Department of Physics and Astronomy, Wayne State University, Detroit, Michigan 48201, USA}
$$$$

\author{Berndt M\"uller}
\affiliation{Department of Physics, Duke University, Durham, North Carolina 27708, USA}
\affiliation{Brookhaven National Laboratory, Upton, NY 11973, USA}
$$$$

\date{\today}
\begin{abstract}

We study elliptic and triangular flow in the collisions of deuteron-gold nuclei at $\sqrt{s_{\rm NN}} = 200$~GeV at RHIC and of proton-lead nuclei at $\sqrt{s_{\rm NN}} = 5.02$~TeV at the LHC, utilizing (3+1)-dimensional ideal hydrodynamics for the dynamic evolution of the fireball and a Monte-Carlo Glauber based energy deposition model for simulating the fluctuating initial conditions. Sizable values of elliptic and triangular flow are obtained for both colliding systems, and the results are consistent with PHENIX, ALICE, ATLAS and CMS measurements. For these studied centralities, we find that the elliptic flow in proton-lead collisions is smaller than deuteron-gold collisions, while the triangular flows are comparable in both colliding systems.
Our results indicate that the observed collective anisotropic flow in deuteron-gold and proton-lead collisions may be obtained from relativistic hydrodynamic evolution of the fireball with initial state fluctuations.

\end{abstract}
\maketitle

\section{Introduction}

Anisotropic flow \cite{Ollitrault:1992bk, Voloshin:2008dg} is one of the most important probes of the hot, dense quark-gluon plasma (QGP) created in ultra-relativistic nuclear collisions, such as those at the Relativistic Heavy-Ion Collider (RHIC) and the Large Hadron Collider (LHC).
Relativistic hydrodynamics has been very successful in describing the dynamical evolution of the hot and dense fireball created in these energetic collisions, especially the observed flow anisotropy for the produced particles \cite{Adams:2003zg, Aamodt:2010pa, ATLAS:2011ah, Chatrchyan:2012ta}.
Being understood as the hydrodynamic response to the almond geometry of the collision zone between two colliding nuclei, elliptic flow $v_2$ has been extensively used to for the quantitative extraction of the transport properties, such as the shear viscosity to entropy ratio $\eta/s$ of the produced dense QCD matter \cite{Luzum:2008cw, Dusling:2007gi, Schenke:2010rr, Song:2010mg}.

Recent studies have demonstrated that the geometric fluctuations in the initial states, such as the positions of nucleons or color charges inside the two colliding nuclei \cite{Miller:2003kd, Broniowski:2007ft, Alver:2008zza, Hirano:2009bd}, may give rise to many rich phenomena, such as finite elliptic shape and flow in the collisions with almost zero impact parameter \cite{Alver:2006wh}, and the presence of odd harmonic moments in the initial geometry and final momentum anisotropy \cite{Alver:2010gr, Alver:2010dn, Petersen:2010cw, Shuryak:2009cy, Staig:2010pn, Qin:2010pf, Lacey:2010hw, Nagle:2010zk, Ma:2010dv, Xu:2010du, Teaney:2010vd, Qiu:2011iv, Bhalerao:2011yg, Floerchinger:2011qf}.
The measurements of the third harmonic flow $v_3$ and other higher harmonic flow available from RHIC and the LHC heavy ion experiments \cite{Adare:2011tg, ATLAS:2011ah} have attracted a lot of attention, and triggered significant effort to investigate the fluctuations of the initial states, their dynamical evolution, and their hydrodynamic responses and manifestation in the final states \cite{Petersen:2010cw, Qin:2010pf, Schenke:2010rr, Teaney:2010vd, Qiu:2011iv, Qin:2011uw, Qiu:2012uy, Pang:2012he}.
One of the goals of such studies to obtain a comprehensive understanding the expansion dynamics and transport properties of the fireball produced in relativistic nuclear collisions.

The proton-nucleus and deuteron-nucleus collisions at ultra-relativistic energies are of great interest as well since they are expected to provide the baseline measurements for heavy-ion collisions \cite{Salgado:2011pf}.
Since the produced QCD matter have smaller sizes compared to heavy ion collisions, one expects weaker collective behavior and anisotropic flow in such colliding systems.
Interestingly, recent experimental measurements have observed clear collective behavior in both proton-lead (p-Pb) collisions at the LHC energies and deuteron-gold (d-Au) collisions at RHIC energies \cite{Abelev:2012ola, Aad:2012gla, Chatrchyan:2013nka, Adare:2013piz}, and even in the proton-proton collisions with high multiplicity at the LHC energies \cite{Khachatryan:2010gv}.
These results have triggered significant interest in finding out the origin of the observed collective behavior in these smaller colliding systems: whether it is an early time effect such as gluon saturation dynamics \cite{Dusling:2012iga} or is generated by late time hydrodynamic expansion of the fireball \cite{Bozek:2011if, Bozek:2013uha, Bzdak:2013zma}.
The answer to this question may significantly affect our understanding of collective behaviors and transport properties of dense QCD matter produced in relativistic heavy-ion collisions.

In this work, we present our study of the anisotropic flow for both deuteron-gold collisions at $\sqrt{s_{\rm NN}} = 200$~GeV at RHIC and proton-lead collisions at $\sqrt{s_{\rm NN}} = 5.02$~TeV at the LHC.
Our study is not focused on the effect of dissipation, i.~e.\ shear viscosity, on the final particle anisotropy -- others have already explored this aspect \cite{Bozek:2011if, Bozek:2013uha} --  but on the question whether the spatial distribution of energy-momentum deposition in these highly asymmetric collision systems is consistent with a collective flow interpretation of the final state anisotropies.
The initial conditions are simulated via a Monte-Carlo Glauber based energy deposition model that we developed in an earlier study of anisotropic flow in heavy-ion collisions \cite{Qin:2010pf}.
Our energy deposition initial condition model has not yet been applied to study the much smaller d-Au and p-Pb collision systems.

In our initial condition model, the multiplicity fluctuations as well as the fluctuations of energy-momentum deposition in the initial collision are both included; the later ingredient has often been neglected.
The inclusion of momentum deposition in the initial states allows us to simulate the pre-equilibrium evolution of the system, during which some initial flow may be developed prior to the application of hydrodynamic evolution.
Here we apply the free-streaming approximation for the dynamics of the pre-equilibrium stage, which has been shown to be a good treatment for early time dynamics even in the strong coupling limit \cite{Balasubramanian:2013rva, Balasubramanian:2013oga}.
For late stage dynamical evolution and collective motion of the produced dense QCD matter, we utilize a (3+1)-dimensional ideal hydrodynamic model from Ref. \cite{Rischke:1995ir, Rischke:1995mt}.
With the same setup, we calculate the elliptic and triangular flow for both d-Au collisions at RHIC and p-Pb collisions at the LHC. We compare our numerical results to available measurements from PHENIX, ALICE, ATLAS and CMS Collaborations.

\section{Setup}

We start with our initial conditions, which are built based on the Monte-Carlo Glauber model \cite{Miller:2003kd}, with the effect of multiplicity fluctuations and initial flow fluctuations implemented as in Ref. \cite{Qin:2010pf}.
The nuclear distribution function inside a nucleus is taken as the Woods-Saxon form,
\begin{align}
\rho_A(r) \propto 1/\{1 + \exp[(r-R)/d]\},
\end{align}
where the parameters $R=6.38~{\rm fm}$, $d=0.535~{\rm fm}$ for a Au nucleus, and $R=6.62~{\rm fm}$, $d=0.546~{\rm fm}$ for a Pb nucleus.
For the deuteron, we employ the Hulthen form of the wave function
\begin{eqnarray}
\phi(r) \propto (e^{-A r} - e^{-B r})/r,
\end{eqnarray}
with the parameters set as $A=0.228{\rm fm}^{-1}$ and $B=1.18{\rm fm}^{-1}$ \cite{Miller:2003kd}.
In Glauber model simulation, one of the important inputs is the inelastic nucleon-nucleon cross section $\sigma_{\rm NN}$, which we take as $42~{\rm mb}$ for $\sqrt{s_{\rm NN}} = 200~{\rm GeV}$ collisions and $67.7~{\rm mb}$ for $\sqrt{s_{\rm NN}} = 5.02~{\rm TeV}$ collisions.

The multiplicity fluctuations in A+B collisions are incorporated by evaluating the following phenomenological equation on an event-by-event basis,
\begin{eqnarray}\label{2comp_n}
N_{AB}  =  \left[\alpha  N_{\rm coll}  + ({1 - \alpha})N_{\rm part}/{2}  \right] N_{\rm NN}.
\end{eqnarray}
Here the event distribution of particle multiplicity $N$ in nucleon-nucleon collisions is taken as a negative binomial distribution,
\begin{align}
P(N, \mu, k) = \frac{\Gamma(N+k)}{\Gamma(N+1) \Gamma(k)}
\frac{(\mu/k)^{N}}{(\mu/k + 1)^{N+k}},
\label{NBD}
\end{align}
where $\mu$ is the mean of the distribution, and $k$ controls the shape of the distribution.
In this application, their values are taken as $\mu = 2.35$, $k=1.9$ for $\sqrt{s_{\rm NN}} = 200~{\rm GeV}$ \cite{Gans, Qin:2010pf} and $\mu= 5.36$, $k=0.9$ for $\sqrt{s_{\rm NN}} = 5.02~{\rm TeV}$ \cite{CMS:2012, Bozek:2013uha}.
The balance factor $\alpha$ in Eq. \ref{2comp_n} controls the relative contributions from participating nucleons and binary collisions, and are usually fixed by fitting to the centrality dependence of particle multiplicity. Here we take $\alpha = 0$ for both d-Au and p-Pb collisions at all centralities.

{We further assign each produced particle a momentum so that one may simulate the pre-equilibrium evolution of the system and the development of initial flow fluctuations during this stage (the free-streaming approximation is used for this work).}
The transverse momenta of the particles are sampled according to the following distribution,
\begin{align}
\frac{dN}{dp_T^2} \propto \frac{1}{(1 + p_T^2/b^2)^c}.
\end{align}
For for d-Au collisions at $\sqrt{s_{\rm NN}} = 200~{\rm GeV}$ the parameters are taken as $b=1.18$, $c=4.09$ \cite{Adler:2003ii}, and for p-Pb collisions at $\sqrt{s_{\rm NN}} = 5.02~{\rm TeV}$ they are $b=1.12$, $c=3.08$ \cite{ALICE:2012mj}.
The rapidity of each produced particle is determined according to the following distribution \cite{Bozek:2013uha},
\begin{eqnarray}
P(\eta) \propto \exp\left[{-\frac{(|\eta| - \eta_0)^2}{2\sigma_\eta^2} \theta(|\eta| - \eta_0)}\right] F(\eta),
\end{eqnarray}
where we set $\eta_0 = 1.9$, $\sigma_\eta=2.0$ for d-Au collisions at RHIC, and $\eta_0=2.5$, $\sigma_\eta=1.4$ for p-Pb collisions at the LHC.
The $F$ function in the above equation depends on the origins of the produced particles.
For particles originating from the participants in deuteron (proton) moving in $+z$ direction, we take it as $F=(1 + \eta/y_b)\theta(y_b + \eta)$, with $y_b$ the beam rapidity, while for particles from the participating nucleons in Au (Pb) nucleus moving in $-z$ direction, $F=(1 - \eta/y_b)\theta(y_b - \eta)$.
For the particles contributed from binary collisions, a symmetric distribution for $P(\eta)$ will be used, i.e., $F=1$.

With the above setup, we obtain the full phase distribution $f(\vec{x}, \vec{p})$ of the system at initial production time.
We may calculate the initial participant eccentricities as follows,
\begin{align}
\epsilon_n = {\sqrt{\langle r_\perp^n \cos(n\phi) \rangle^2 + \langle r_\perp^n
\sin(n\phi) \rangle^2}} / {\langle
r_\perp^n \rangle},
\end{align}
where $r_\perp = \sqrt{x^2+y^2}$, and $\phi=\arctan(y/x)$ are polar coordinates in the transverse plane.
The participant plane angle $\Phi_n$ with respect to the reaction plane can be found through the following formula,
\begin{align}\label{Phin}
\Phi_n = \frac{1}{n} \arctan \frac{\langle r_\perp^n \sin(n\phi)\rangle}
{\langle r_\perp^n \cos(n\phi) \rangle}.
\end{align}
Note that the function $\arctan(y/x)$ gives the angle between the positive $x$-axis and the point $(x,y)$ in the transverse plane.


\begin{figure}[tbh]
\includegraphics[width=0.91\linewidth]{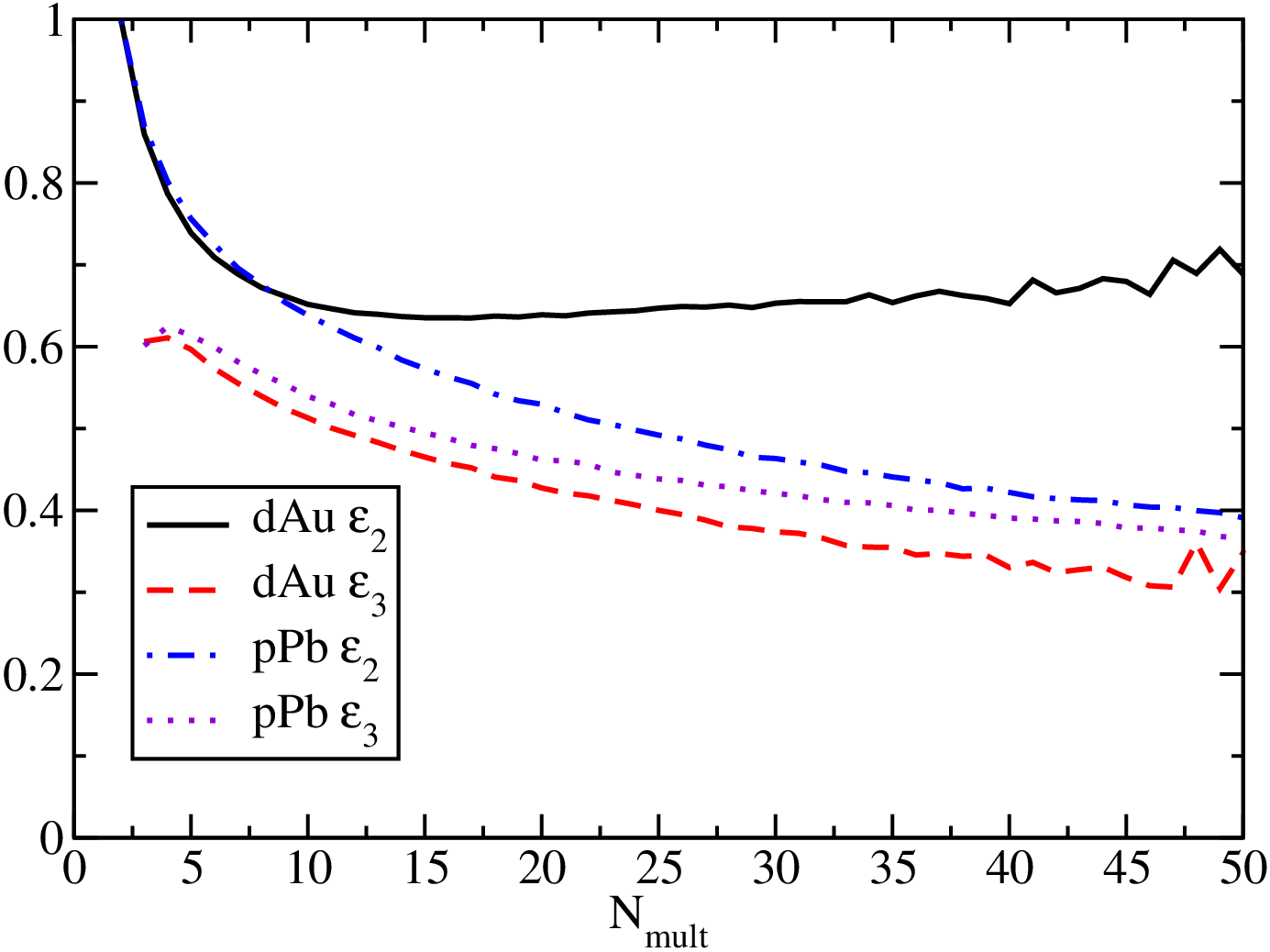}
 \caption{(Color online) The initial eccentricities $\epsilon_2$ and $\epsilon_3$ as a function of $N_{\rm mult}$, for d-Au collisions at $\sqrt{s_{\rm NN}}=200$~GeV and for p-Pb collisions at $\sqrt{s_{\rm NN}}=5.02$~TeV. }
\label{e2e3_nmult}
\end{figure}

\begin{figure}[tbh]
\includegraphics[width=0.91\linewidth]{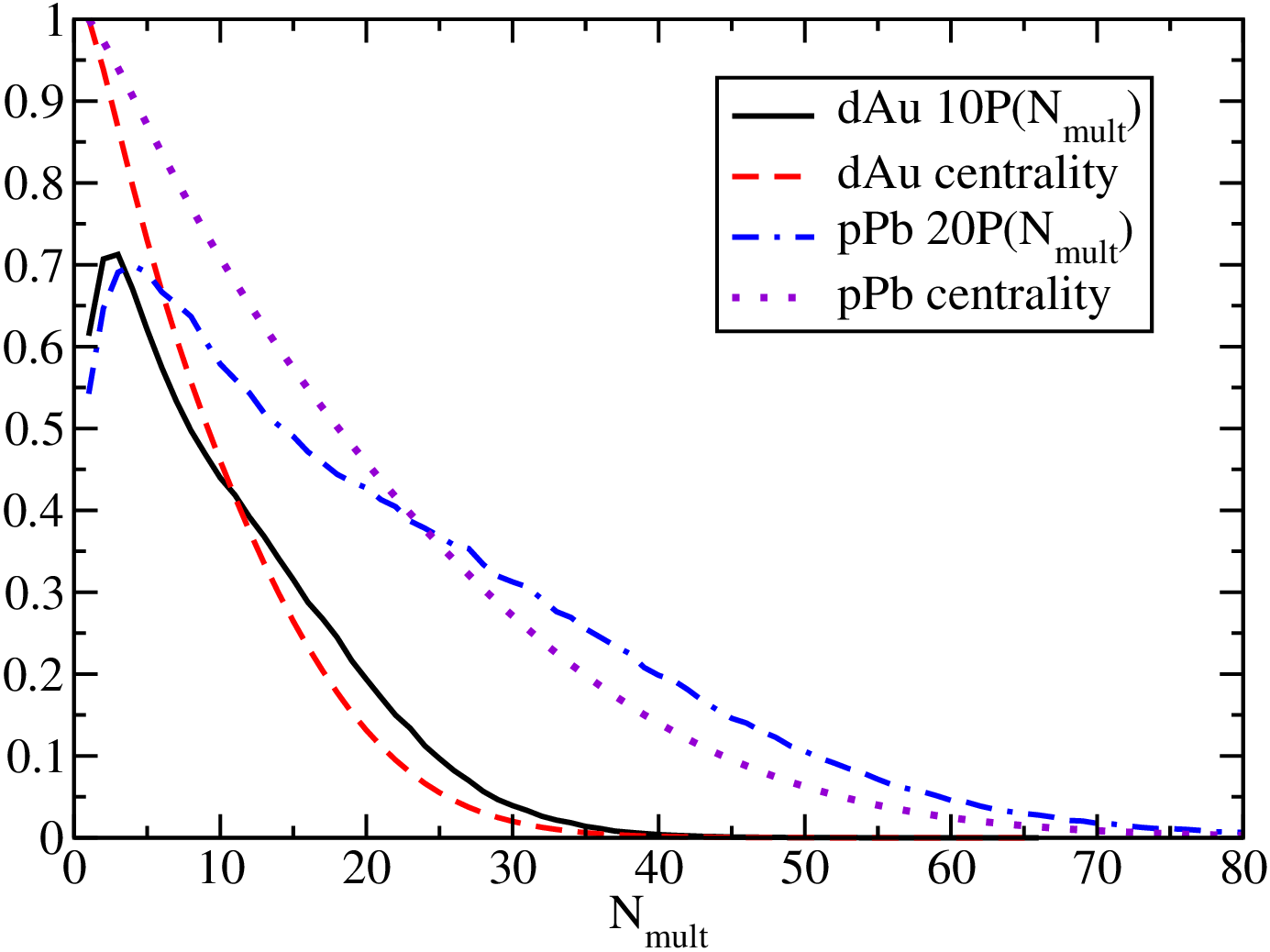}
 \caption{(Color online) The probability distribution of particle number $P(N_{\rm })$ and the centrality as a function of $N_{\rm mult}$, for d-Au collisions at $\sqrt{s_{\rm NN}}=200$~GeV and p-Pb collisions at $\sqrt{s_{\rm NN}}=5.02$~TeV.}
\label{P_nmult}
\end{figure}


The calculated initial geometry anisotropy parameters for both d-Au collisions at $\sqrt{s_{\rm NN}}=200$~GeV at RHIC and for p-Pb collisions at $\sqrt{s_{\rm NN}}=5.02$~TeV at the LHC are shown in Fig. \ref{e2e3_nmult}, where the event average values of the participant eccentricities $\epsilon_2$ and $\epsilon_3$ are plotted as a function of the initial particle multiplicity $N_{\rm mult}$ of the collisions.
One may observe sizable values for the initial anisotropy parameters for both colliding systems in the presence of initial state fluctuations.
While the values of the eccentricity $\epsilon_2$ for p-Pb collisions is smaller than these in d-Au collisions, the values of $\epsilon_3$ are comparable for both colliding systems.
{This is a natural result of our Glauber-based initial condition model since d-Au collision systems appear to be a dumbbell shape, and thus produce larger values of $\epsilon_2$ than p-Pb collision systems.}

{To compare with final experimental data, we here use the initial particle multiplicity $N_{\rm mult}$ to determine the centrality of the collisions.
In Fig. \ref{P_nmult}, we show the probability distribution of particle multiplicity $N_{\rm mult}$ for d-Au collisions at $\sqrt{s_{\rm NN}} = 200$~GeV and p-Pb collision at $\sqrt{s_{\rm NN}} = 5.02$~TeV, respectively. Using these distributions, one may determine the centrality which are shown in the same figures.
We note that other centrality determination methods may be used, e.g., the participant nucleon number $N_{\rm part}$ \cite{Bozek:2011if}.
In the last part, we will compare the results using three different methods: particle multiplicity, participant number, and deposited transverse energies $E_T$, to determine the centrality.
}

The above initial conditions will serve as the inputs for relativistic hydrodynamics simulation of the produced fireball.
In this work, we follow the common practice to assume a sudden thermalization of the system at the time $t_0$ which we take to be $0.7~{\rm fm}/c$ for both d-Au collisions at RHIC and p-Pb collisions at the LHC.
Prior to this time, the system is taken to be free-streaming to include some of the pre-equilibrium effects, such as the development of initial flow, the decreasing of geometric anisotropy and the mixing of anisotropy of different orders \cite{Qin:2010pf}.

The energy-momentum tensor is calculated from the phase space distribution $f({\bf x}, {\bf p})$,
\begin{align}
T^{0 \nu }(x) = \int \frac{d^3p}{E} p^{0}p^{\nu } f\left({\bf x}, {\bf p}\right).
\end{align}
For our discretized phase space distribution $f({\bf x}, {\bf p}) = \sum_{i} \delta({\bf x} - {\bf x}_i) \delta({\bf p} - {\bf p}_i)$, the momentum integration $\int d^3 p$ in the above turns into the sums over all particles.
The spatial part of the discretized distribution is smeared with a Gaussian function,
\begin{eqnarray}
\hspace{-24pt} \delta({\bf x} - {\bf x}_i) \rightarrow \frac{
\exp\left[-\frac{(x-x_i)^2+(y-y_i)^2}{2\sigma_{xy}^2}\right]
}{{2\pi\sigma_{xy}^2}}
 \frac{\exp\left[-\frac{(z-z_i)^2}{2\sigma_{z}^2}\right]}{\sqrt{2\pi\sigma_{z}^2
}}.
\end{eqnarray}
Such treatment is necessary for the use of hydrodynamic simulation. In this application, the widths $\sigma_{xy}$, $\sigma_z$ are fixed to be the same as the starting time $t_0$ of the hydrodynamic simulation.

After obtaining the energy momentum tensor as described above, we may start the ideal hydrodynamical simulation,
\begin{align}
\partial_\mu T^{\mu\nu}(x) = 0.
\end{align}
With the assumption of a sudden thermalization, we take the calculated energy and momentum densities $(T^{00}, T^{0i})$ as the inputs to an ideal hydrodynamical evolution code \cite{Rischke:1995ir, Rischke:1995mt} with the use of a lattice equation of state \cite{Steinheimer:2009nn, Steinheimer:2009hd} for the hot and dense matter created in d-Au and p-Pb collisions.
The off-equilibrium part of the energy momentum tensor is neglected in this application, and may be included if one extends to viscous hydrodynamical simulation.
{We assume that the net baryon number density vanishes and set $\mu_B = 0$.}
When the matter is diluted in the late stage, the production of particles at the end of the hydrodynamic evolution is treated as a gradual freeze-out on an approximated iso-eigentime hyper-surface according to the Cooper-Frye prescription \cite{Li:2008qm, Steinheimer:2009nn}.

In this work, we do not perform a complete event-by-event hydrodynamic simulation for each fluctuating initial condition for the purpose of saving computing time.
Rather, for each centrality bin we first perform an average over $5000$ events of initial conditions with a rotation of each event by an angle of $\Phi_2$ when calculating elliptic flow $v_2$, and a rotation of $\Phi_3$ for calculating triangular flow $v_3$.
Utilizing such event-averaged initial condition profiles, we then perform one-shot hydro simulations. For the particle production at freeze-out, we generate $2000$ events for d-Au collisions and for p-Pb collisions.
We do not perform the simulation of hadronic rescattering in the dilute hadron gas stage and the resonance decays; they should not have much influence on the results for the charged particle flow coefficients \cite{Petersen:2010md}.
However, the resonance decays will increase the final particle multiplicity, which has been shown to be about a factor of two for charged hadrons \cite{Qiu:2012tm}.
Taking into account such a factor of two effect, we tune the overall normalization of our initial conditions to obtain the final charged hadron multiplicities for $|\eta|< 2.5$, which are about $90$ for 0-20\% d-Au collisions at RHIC and about $150$ for 0-2\% p-Pb collisions at the LHC.

Based on the above simulations, we may analyze the anisotropic flow for d-Au collisions at $\sqrt{s_{NN}}=200$~GeV at RHIC and for p-Pb collisions at $\sqrt{s_{NN}}=5.02$~TeV at the LHC.
{The flow coefficients $v_n$ are calculated using the two-particle correlation method: $v_n=v_{n,n}/\sqrt{|v_{n,n}|}$, with $v_{n,n} = \langle \cos[n(\psi_1 - \psi_2)] \rangle$, where $\psi_1$ and $\psi_2$ are the azimuthal angles of a pair of particles.}

\section{Results}


\begin{figure}[tbh]
\includegraphics[width=0.91\linewidth]{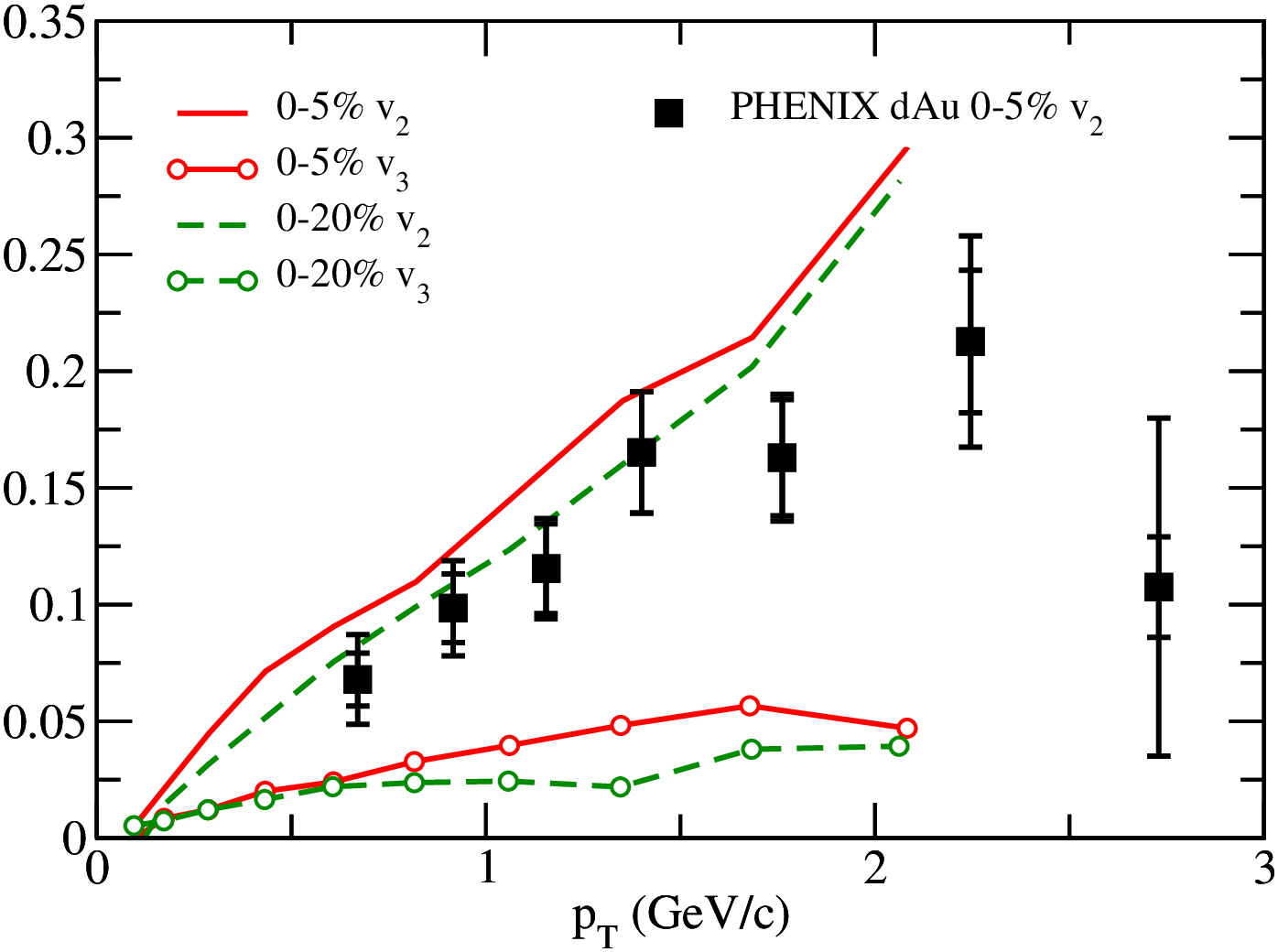}
 \caption{(Color online) Elliptic and triangular flow as a function of $p_T$ for d-Au collisions at $\sqrt{s_{\rm NN}}=200$~GeV at RHIC for 0-5\% and 0-20\% centralities.}
\label{dAu}
\end{figure}

\begin{figure}[tbh]
\includegraphics[width=0.91\linewidth]{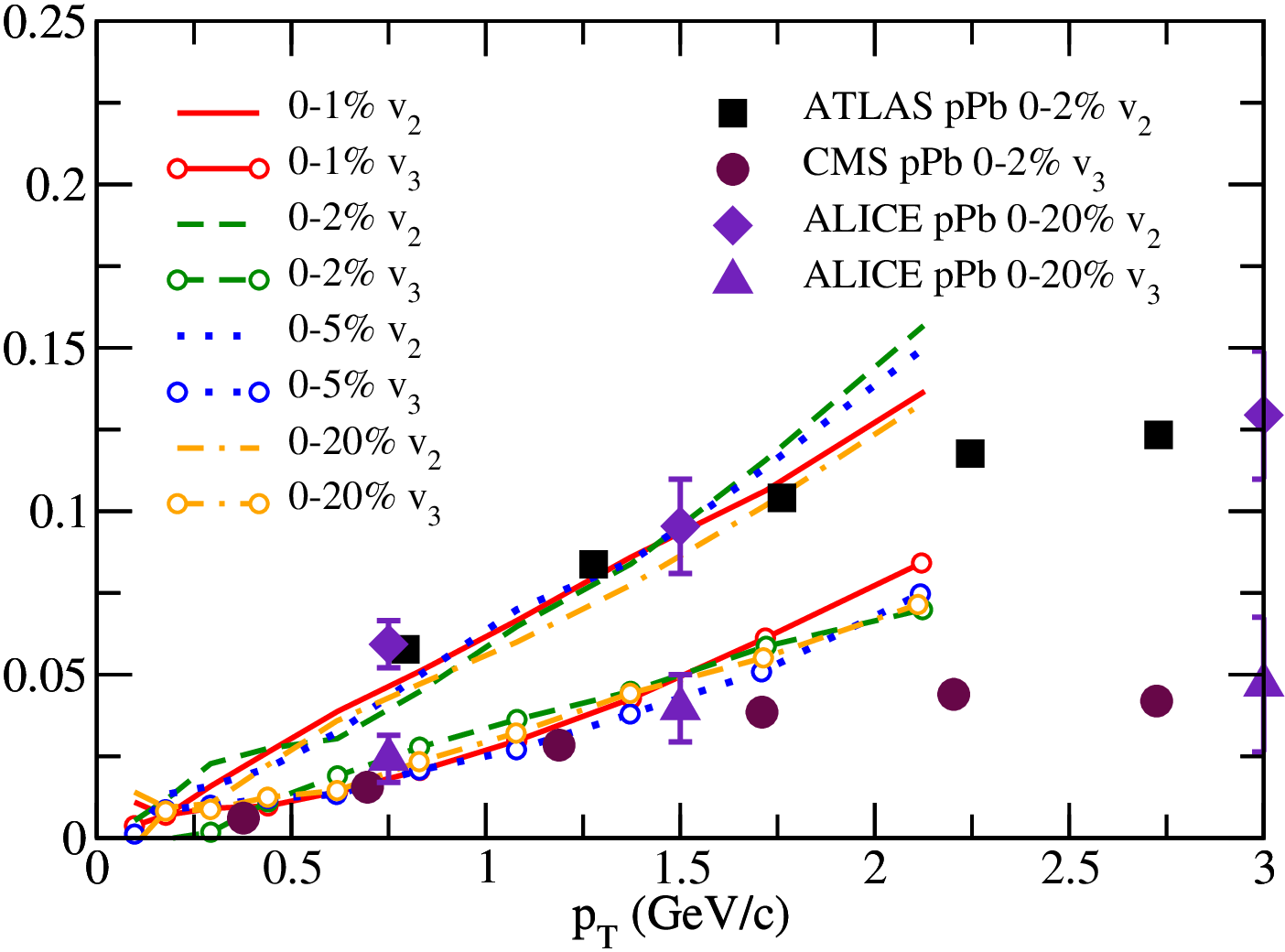}
 \caption{(Color online)  Elliptic and triangular flow as a function $p_T$ for p-Pb collisions at $\sqrt{s_{\rm NN}}=5.02$~TeV at the LHC for 0-1\%, 0-2\%, 0-5\% and 0-20\% centralities.}
\label{pPb}
\end{figure}


Now we present the results for the final anisotropic flow for d-Au collisions at $\sqrt{s_{\rm NN}}=200$~GeV at RHIC and for p-Pb collisions at $\sqrt{s_{\rm NN}}=5.02$~TeV at the LHC.
Before comparing with the experimental data which use two-particle correlation methods, we multiply our $v_n$ results by a factor $2/\sqrt{\pi}$ to include the flow fluctuations due to initial geometry fluctuations \cite{Broniowski:2007ft, Voloshin:2007pc}.
Such treatment is good when the initial fluctuations dominate the anistropic flow (e.g., $v_3$ or very central collisions), and may overestimate the flow fluctuations as the initial geometry becomes important (e.g., $v_2$ in d-Au collisions and less central p-Pb collisions).

In Fig. \ref{dAu}, the elliptic flow $v_2$ and triangular flow $v_3$ for d-Au collisions are shown as a function of transverse momentum $p_T$.
The results for two different centralities, 0-5\% and 0-20\% are shown for comparison.
One may observe sizable and comparable values of elliptic flow for both centralities. The elliptic flow result for 0-5\% centrality class is consistent with measurement from PHENIX Collaboration \cite{Adare:2013piz}.
The triangular flows for both centralities have similar magnitude, and are smaller than elliptic flow.

In Fig. \ref{pPb}, we show the elliptic and triangular flows as a function $p_T$ for p-Pb collisions for four different centralities, 0-1\%, 0-2\%, 0-5\% and 0-20\%.
The elliptic and triangular flows for these centralities are comparable, and the triangular flows $v_3$ are smaller than elliptic flow.
The results of $v_2$ and $v_3$ are comparable with the data measured by ALICE, ATLAS and CMS Collaborations \cite{Abelev:2012ola, Aad:2012gla, Chatrchyan:2013nka}. When comparing to d-Au collisions as shown in Fig. \ref{dAu}, we observe that while the elliptic flow $v_2$ for p-Pb collisions at the LHC is smaller for these studied centralities, the triangular flows from both colliding systems are comparable.
This is quite similar to the results for the initial geometry anisotropy parameter as shown in Fig. \ref{e2e3_nmult}.
This might suggest that much of the final observed anisotropic flow in p-Pb collisions and d-Au collisions may be developed from the geometric anisotropy and fluctuations of the initial states via late hydrodynamic evolution.
{We note that our calculation is based on ideal hydrodynamics; the use of viscous hydrodynamics would suppress the development of both anisotropic flows. The suppression will be more for triangular flow than elliptic flow.
}


\begin{figure}[tbh]
\includegraphics[width=0.91\linewidth]{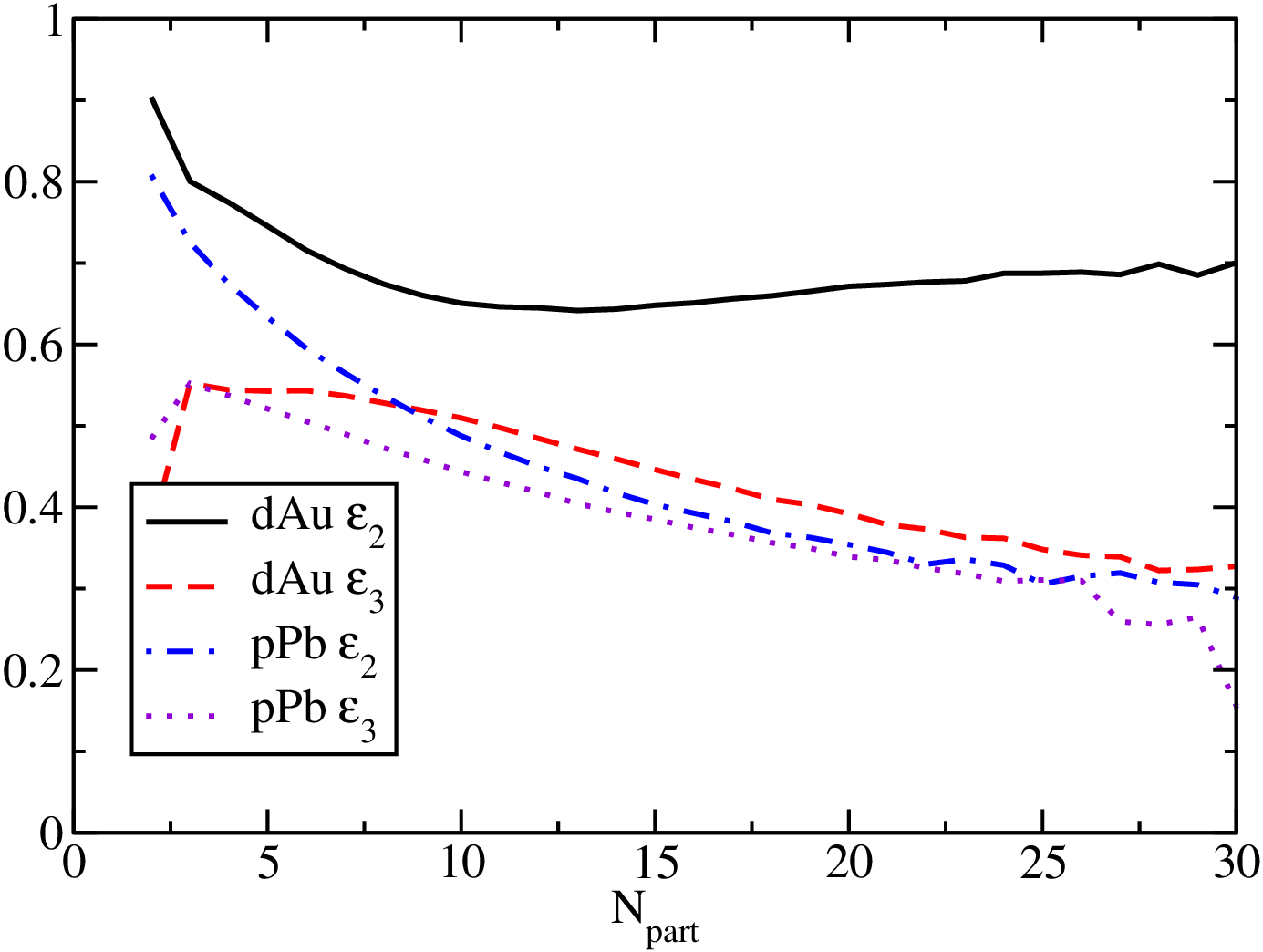}
 \caption{(Color online) The initial eccentricities $\epsilon_2$ and $\epsilon_3$ as a function of $N_{\rm part}$ for d-Au collisions at $\sqrt{s_{\rm NN}}=200$~GeV, and for p-Pb collisions at $\sqrt{s_{\rm NN}}=5.02$~TeV. }
\label{e2e3_npart}
\end{figure}

\begin{figure}[tbh]
\includegraphics[width=0.91\linewidth]{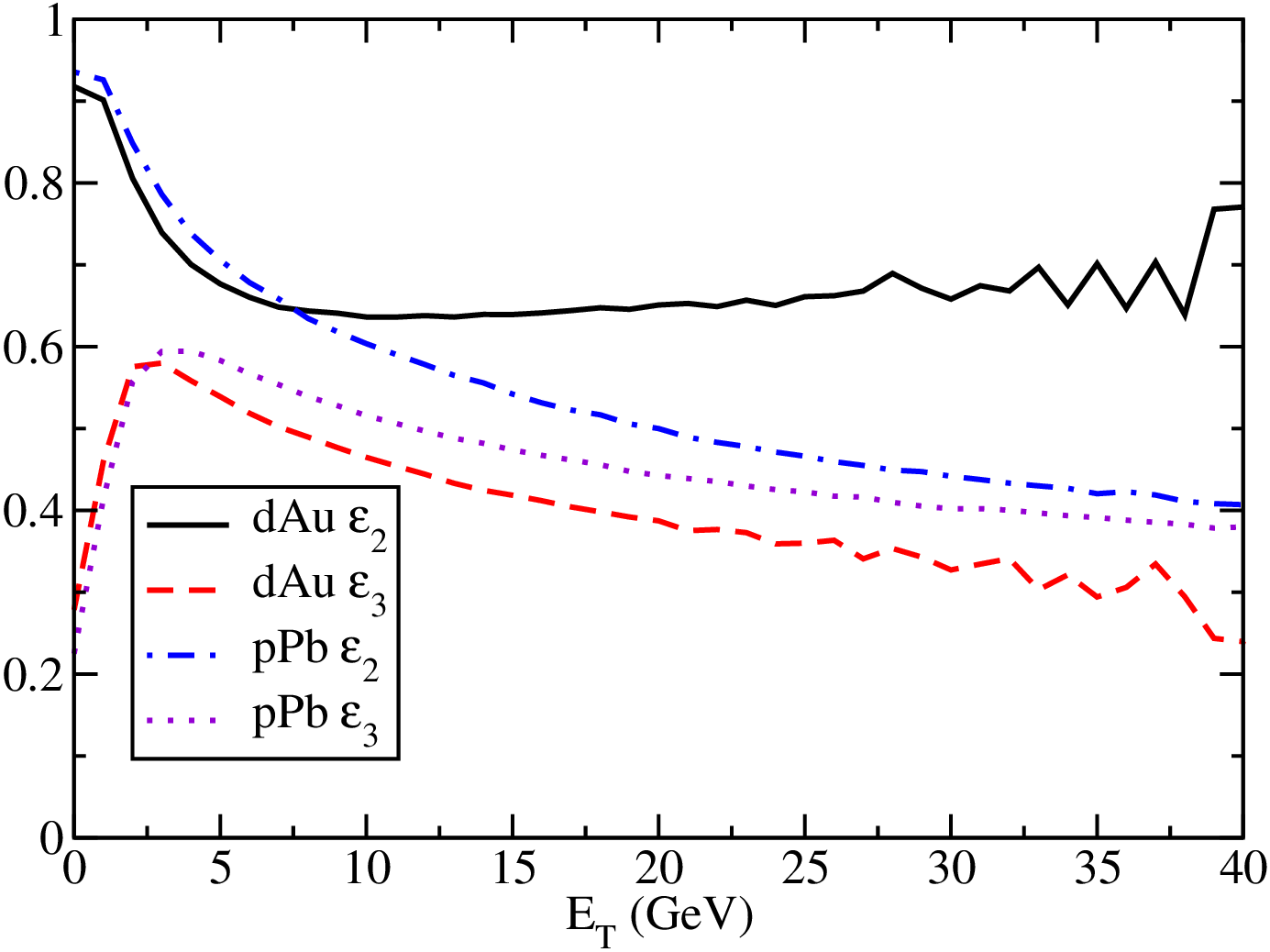}
 \caption{(Color online) The initial eccentricities $\epsilon_2$ and $\epsilon_3$ as a function $E_{T}$ for d-Au collisions at $\sqrt{s_{\rm NN}}=200$~GeV, and for p-Pb collisions at $\sqrt{s_{\rm NN}}=5.02$~TeV. }
\label{e2e3_et}
\end{figure}


\begin{figure}[tbh]
\includegraphics[width=0.91\linewidth]{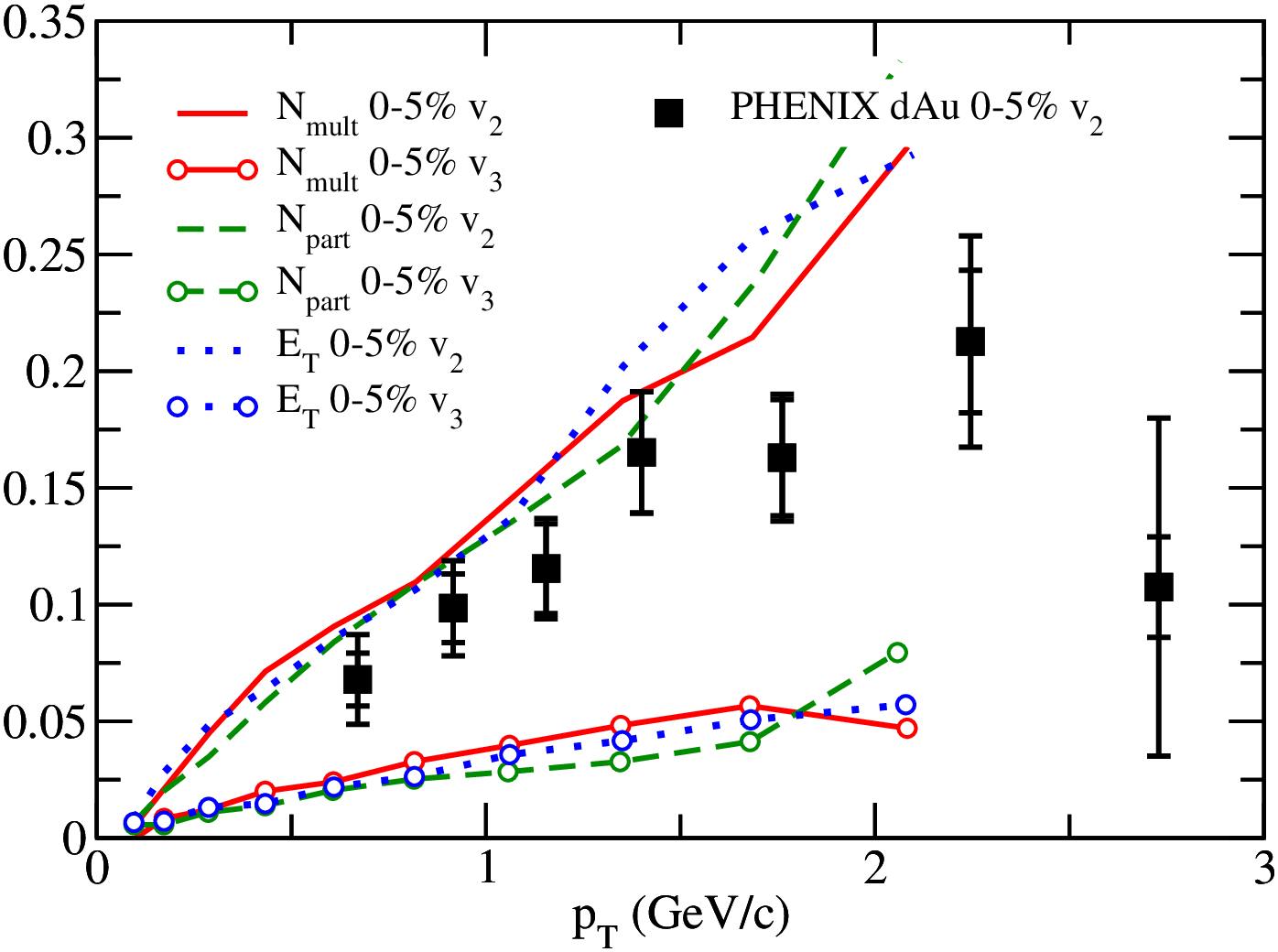}
 \caption{(Color online) $v_2$ and $v_3$ as a function of $p_T$ for d-Au collisions at $\sqrt{s_{\rm NN}}=200$~GeV at RHIC for 0-5\% centrality, using different centrality determination methods.}
\label{v23_dAu}
\end{figure}

\begin{figure}[tbh]
\includegraphics[width=0.91\linewidth]{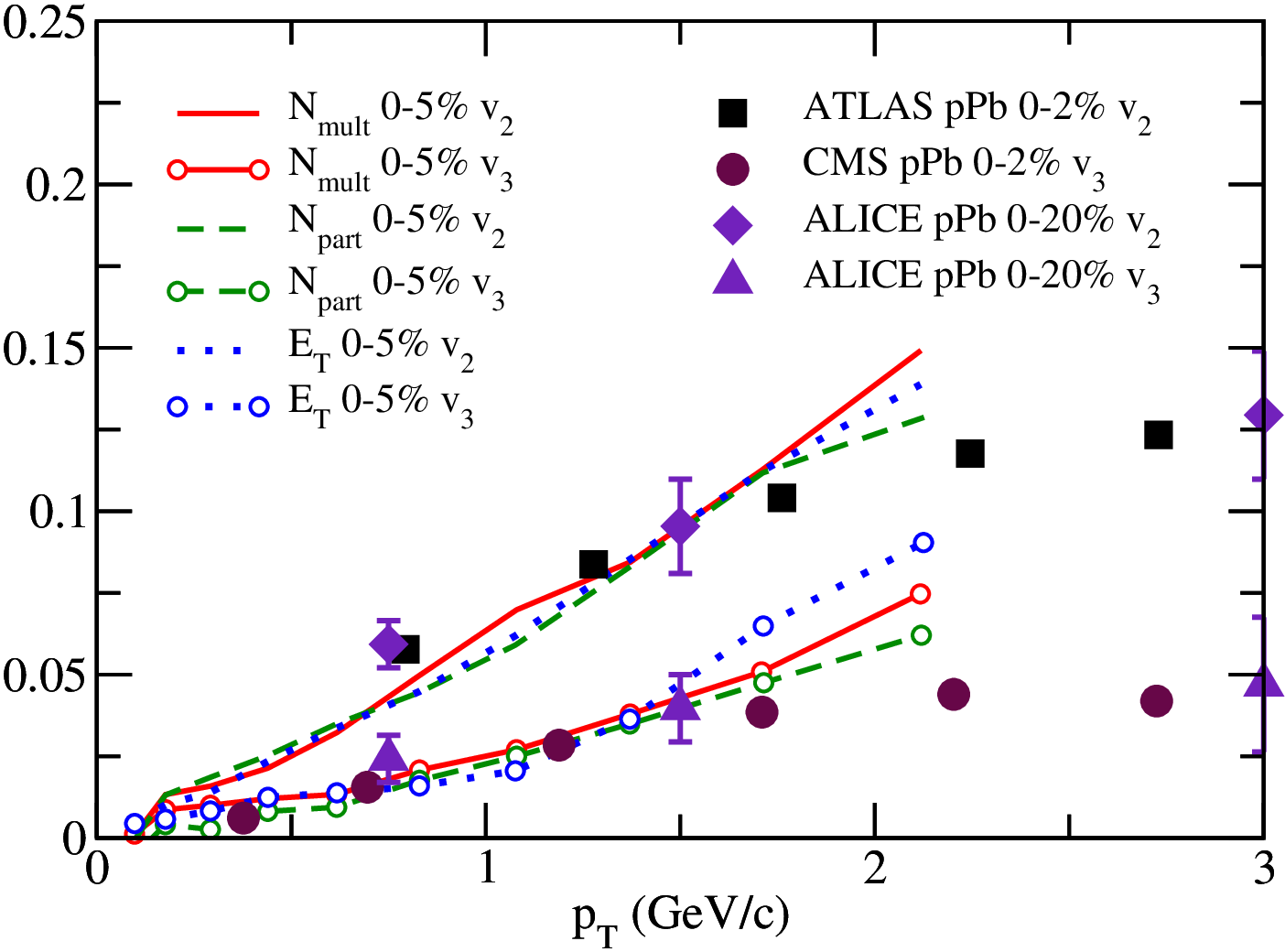}
 \caption{(Color online)  $v_2$ and $v_3$ as a function $p_T$ for p-Pb collisions at $\sqrt{s_{\rm NN}}=5.02$~TeV at the LHC for 0-5\% centrality, using different centrality determination methods.}
\label{v23_pPb}
\end{figure}


{In the above results, we have used the initial particle multiplicity to determine the collision centrality to be compared with the experimental data.
The use of participant number $N_{\rm part}$ is another popular method for centrality determination in hydrodynamic simulation and flow calculation \cite{Bozek:2011if}.
In our Glauber-based model, we have included both multiplicity fluctuations and initial flow fluctuations; this allow us to determine the collision centrality using the initial transverse energies $E_T$ of the collisions.
We do not compare the calculation using impact parameter for centrality determination; such method has been proved to be not useful in event-by-event hydrodynamic simulations (see e.g., Ref. \cite{Bozek:2011if}).
We also note that although different quantities are used in our simulation for the centrality determination, the multiplicity fluctuations as well as initial flow fluctuations are still included in the initial conditions.
For different methods, we tune the overall normalization factor of our initial conditions to get the same final state charged hadron multiplicities.
}

{Now we compare the results using these three different centrality determination methods.
In Fig. \ref{e2e3_npart} and \ref{e2e3_et}, we show the initial geometry anisotropy parameters $\epsilon_2$ and $\epsilon_3$ for both d-Au collisions at $\sqrt{s_{\rm NN}}=200$~GeV at RHIC and for p-Pb collisions at $\sqrt{s_{\rm NN}}=5.02$~TeV at the LHC, as a function of participant number and transverse energies.
One may see that the results are very similar to Fig. \ref{e2e3_nmult}, in which the anisotropy parameters are plotted as a function of initial multiplicity.
We obtain larger values of $\epsilon_2$ for d-Au collisions than for p-Pb collisions from our Glauber-based initial conditions; the sizes of $\epsilon_3$ are comparable for both colliding systems.
}

{In Fig. \ref{v23_dAu} and \ref{v23_pPb}, we show the results of elliptic and triangular flow as a function $p_T$ for both d-Au collisions at $\sqrt{s_{\rm NN}}=200$~GeV at RHIC and for p-Pb collisions at $\sqrt{s_{\rm NN}}=5.02$~TeV at the LHC.
To compare the results from three different centrality determination methods: the initial multiplicity, the participant number and initial transverse energies, we only show the results for 0-5\% centrality.
We observe that three different methods give similar magnitudes of $v_2$ and $v_3$ for both d-Au and p-Pb collisions (0-5\% centrality).
This may be understood as resulting from the weak centrality dependence of eccentricities as shown in Fig. (\ref{e2e3_nmult}, \ref{e2e3_npart}, \ref{e2e3_et}) in most central d-Au and p-Pb collisions.
}

\section{Summary}

In summary, we have performed a study of elliptic and triangular flow for deuteron-gold collisions at $\sqrt{s_{\rm NN}} = 200$~GeV at RHIC and proton-lead collisions at $\sqrt{s_{\rm NN}} = 5.02$~TeV at the LHC. The initial conditions of the produced fireball are simulated by a Glauber-based energy deposition model with the incorporation of both multiplicity fluctuations and initial flow fluctuations. The dynamic evolution and the collective motion of the fireball are simulated with a (3+1)-dimensional ideal hydrodynamics model.
We have presented our numerical results for the final anisotropic flow, and obtained sizable values for elliptic flow and triangular flow in both colliding systems.
These results are consistent with the available measurements from PHENIX, ALICE, ATLAS and CMS Collaborations.
Some of the further directions include the comparison of different initial conditions, the investigation of viscosity and other effects, and the use of complete event-by-event hydrodynamic simulations.
These investigations will be helpful for more comprehensive understanding of initial state fluctuations, the expansion dynamics and transport properties of the produced fireball, and the origin of anisotropic flow in relativistic nuclear collisions.

\vspace{1mm}
\section{Acknowledgments}
We thank D. Rischke for providing the three-dimensional relativistic hydrodynamics code and H. Petersen for helpful discussions.
This work was supported in part by the Natural Science Foundation of China under grant no 11375072 and the U. S. Department of Energy under grant DE-FG02-05ER41367.


\bibliographystyle{h-physrev5}
\bibliography{GYQ_refs}
\end{document}